\documentstyle[aps,epsf,preprint]{revtex}
\begin{document}
\draft
\preprint{\vbox{
\hbox{UT-847}
}}  
\title{End-point of the Electroweak Phase Transition using the
auxiliary mass method\\}
\author {Kenzo {\sc Ogure}
and Joe {\sc Sato}$^{*}$}
\address{Institute for Cosmic Ray Research, University of Tokyo, 
Tanashi, Tokyo 188-8502, Japan
\\
$^*$Department of Physics, School of Science, University of Tokyo,
Tokyo 113, Japan}
\maketitle
\begin{abstract}
We study the end-point of the Electroweak phase transition using the
auxiliary mass method.  The end point is $m_H\sim40$ (GeV) in the case
$m_t=0$ (GeV) and strongly depends on the top quark mass. A first
order phase transition disappears at $m_t\sim 160$ (GeV). 
The renormalization effect of the top quark is significant.
\end{abstract}

The Electroweak phase transition is one of the most important phase
transitions at the early universe since it may account for the baryon
number of the present universe\cite{Kuz}.  This phase transition was
first investigated using the perturbation theory of the finite
temperature field theory and predicts the first order phase transition
from an effective potential\cite{Car,Arn}.  The perturbation theory,
however, has difficulty due to an infrared divergence caused by light
Bosons and cannot give reliable results in the case where the Higgs
Boson mass $m_H$, is comparable to or greater than the Weak Boson mass.
Lattice Monte Carlo simulations, therefore, become the most powerful
method and are still used to investigate details of the phase
transition\cite{Rum,Cas,Ilg,Kar,Aok}.  According to these results,
the Electroweak phase transition is of the first order if $m_H$ is less
than an end-point $m_{H,c}\sim70$ (GeV).  It turns to be of the second
order just on the end-point.  Beyond the end-point, we have no phase
transition, which means any observable quantities do not have
discontinuities.  As far as we know, three other non-perturbative
methods predict the existence of the end-point\cite{Buc,Tet,Hub}.  The
end-point is determined below 100 GeV by these three methods.

The auxiliary-mass method is a new method to avoid the infrared
divergence at a finite temperature T\cite{Dru,Ina,Ogu,Ogu2}.  This
method is based on a simple idea as follows.  We first add a large
auxiliary mass to light Bosons, which cause the infrared divergence,
and calculate an effective potential at the finite temperature.  Due
to the auxiliary mass, the effective potential is reliable at any
temperature.  We next extrapolate this effective potential to the true
mass by integrating an evolution equation, which we show later.  We
applied this method to the $Z_2$-invariant scalar model and the
$O(N)$-invariant scalar model, and obtained satisfactory
results\cite{Ina,Ogu,Ogu2}.

We apply the method to the Standard Model and investigate the
Electroweak phase transition in the present paper.  We add an
auxiliary mass $M\gtrsim T$ only to the Higgs Boson, which becomes
very light owing to a cancellation between its negative tree mass and
positive thermal mass for small field expectation values around the
critical temperature.  We notice that the infrared divergence from the
Higgs Boson is always serious if the phase transition is of the second
order or of the weakly first order\cite{Arn}.  In the standard model,
transverse modes of the gauge fields also have small masses at small
field expectation values since they do not have the thermal mass at
one loop order.  It is however, expected that they do have a
thermal mass ($\sim g^2 T$) at the two loop order\cite{Kar,Buc,Ebe}.
Here, $g$ is a gauge coupling constant.  If so, the loop expansion
parameter\cite{Arn} is $\frac{g^2 T}{M_G}\lesssim 1$, even if the field
expectation value is zero. Here, $M_G$ is the mass of the gauge Boson,
which is a sum of a zero-temperature mass and a thermal mass.  We
assume that this actually occurs and the infrared divergence from the
gauge Bosons is not serious.  Since this small thermal mass for the
transverse modes will bring only a slight change to a one-loop
effective potential, we use the one-loop effective potential without
this small mass for the transverse modes.

An effective potential is then calculated as follows in the Landau
gauge\cite{Car,Arn},
\begin{eqnarray}
V(M^2)
&=&
\frac{M^2}{2}\phi^2+\frac{\lambda}{4!}\phi^4
+f_{BT}(m_H^2(\phi))
+3f_{BT}(m_{NG}^2(\phi))\nonumber\\
&&+4f_{BT}(M_W^2(\phi))+4f_{G0}(M_W^2(\phi))\nonumber\\
&&+2f_{BT}(M_{WL}^2(\phi))+2f_{G0}(M_{WL}^2(\phi)) \nonumber\\
&&+2f_{BT}(M_Z^2(\phi))+2f_{G0}(M_Z^2(\phi))\nonumber\\
&&+f_{BT}(M_{ZL}^2(\phi))+f_{G0}(M_{ZL}^2(\phi))\nonumber\\
&&+f_{BT}(M_{\gamma L}^2(\phi))+f_{G0}(M_{\gamma L}^2(\phi))\nonumber\\
&&+12f_{FT}(m_t^2(\phi))+12f_{F0}(m_t^2(\phi))
    \label{ini}
\end{eqnarray}
here,
\begin{eqnarray}
m_H^2(\phi)&=&M^2+\frac{\lambda}{2}\phi^2,\ 
m_{NG}^2(\phi)=M^2+\frac{\lambda}{6}\phi^2,\nonumber\\
M_W^2(\phi)&=&\frac{g_2^2\phi^2}{4},\ 
M_{WL}^2(\phi)=\frac{g_2^2\phi^2}{4}+\frac{11g_2^2T^2}{6},\nonumber\\
M_Z^2(\phi)&=&\frac{(g_2^2+g_1^2)\phi^2}{4},\ 
m_t(\phi)=\frac{g_Y^2 \phi^2}{2}\label{mass}\nonumber
\end{eqnarray}
\begin{eqnarray}
 \left(
  \begin{array}{cc}
      M_{ZL}^2&0\\
      0&M_{\gamma L}^2
      \end{array}
    \right) 
&=&
{\bf T^{\dagger}}
 \left(
  \begin{array}{cc}
      \frac{g_2^2 \phi^2}{4}+\frac{11g_2^2 T^2}{6}&-\frac{g_1 g_2 \phi^2}{4}\\
      -\frac{g_1 g_2 \phi^2}{4}&\frac{g_1^2 \phi^2}{4}+\frac{11g_2^2 T^2}{6}
      \end{array}
    \right) 
    {\bf T}\nonumber
\end{eqnarray}
\begin{eqnarray}
    \label{func}
    f_{BT}(m^2)&=&\frac{T}{2 \pi^2}\int_0^{\infty}dk \ k^2 
    \log{\{1-\exp{(-\frac{\sqrt{k^2+m^2}}{T})}\} }\nonumber\\
    f_{FT}(m^2)&=&\frac{T}{2 \pi^2}\int_0^{\infty}dk \ k^2 
    \log{\{1+\exp{(-\frac{\sqrt{k^2+m^2}}{T})}\} }\nonumber\\
    f_{G0}(m^2)&=&\frac{m^4}{64\pi^2}
    \{\log{(\frac{m^2}{\bar\mu^2})}-\frac{5}{6}\}\nonumber\\
    f_{F0}(m^2)&=&-\frac{m^4}{64\pi^2}
    \{\log{(\frac{m^2}{\bar\mu^2})}-\frac{3}{2}\}.\nonumber
\end{eqnarray}
In the above equations, $\lambda$, $g_2$, $g_1$ and $g_Y$ are coupling
constants for the Higgs Boson, SU(2) gauge field, U(1) gauge field and
top Yukawa respectively.  The matrix ${\bf T}$ is orthogonal and
diagonalizes the mass matrix for the Z Boson and photon at finite
temperature.  We renormalized the effective potential using the
$\overline{MS}$ scheme with a renormalization scale $\bar\mu$.  A
zero-temperature contribution from the Higgs Boson is neglected since
it is small in the mass region we consider.  The ring diagrams are added
only to the Weak Bosons and the Z-Boson since the Higgs Bosons have
auxiliary large mass and do not need the resummation.  We then
extrapolate this effective potential at the auxiliary mass squared
$M^2$ to that of the true mass squared $-\nu^2$ using an evolution
equation.  Since we add the auxiliary mass only to the Higgs Boson,
the evolution equation is same as that for O(4)-invariant scalar
model, which was constructed in\footnote{ We neglected the momentum
dependence of a full self-energy in Ref.\cite{Ogu}.  This corresponds
to the local potential approximation of the systematic derivative
expansion of the effective action. } \cite{Ogu2},
\begin{eqnarray}
    \frac{\partial V}{\partial m^{2}}&=&
     \frac{1}{2}\bar{\phi}^{2}+\frac{1}{4\pi^{2}}
     \int^{\infty}_{0}dk \frac{k^2}{ \sqrt{k^2
     +\frac{\partial^{2}V}{\partial\bar\phi^{2}}}}
     \frac{1}{ e^{\frac{1}{T}\sqrt{k^2
     +\frac{\partial^{2}V}{\partial\bar\phi^{2}}}}-1}\nonumber\\
     &&+\frac{3}{4\pi^{2}}
     \int^{\infty}_{0}dk \frac{k^2}{ \sqrt{k^2
     +\frac{1}{\bar\phi}\frac{\partial V}{\partial\bar\phi}}}
     \frac{1}{ e^{\frac{1}{T}\sqrt{k^2
     +\frac{1}{\bar\phi}
     \frac{\partial V}{\partial\bar\phi}}}-1}.
    \label{evo}
\end{eqnarray}
A non-perturbative effective potential free from the infrared divergence
can be obtained by solving the evolution equation (\ref{evo}) with an
initial condition Eq.(\ref{ini}) numerically.

Before showing our numerical results, we relate the parameters
$\nu^2$, $\lambda$, $g_2$, $g_1$ and $g_Y$ to physical quantities at
the zero-temperature\cite{Arn},
\begin{eqnarray}
    \label{para}
\lambda
&=&
\frac{3m_{H0}^2}{\phi_0^2}-\frac{3}{32\pi^2}
\left[
\frac{3}{2}g_2^4
\left\{\log{\left(\frac{M_{W0}^2}{\bar\mu^2}\right)+\frac{2}{3}}
\right\}\right.
\nonumber\\
&&+
\frac{3}{4}\left(g_1^2+g_2^2\right)^2
\left\{\log{\left(\frac{M_{Z0}^2}{\bar\mu^2}\right)}+\frac{2}{3}
\right\}\nonumber\\
&&-\left.12g_Y^2\log{\left(\frac{m_{t0}^2}{\bar\mu^2}\right)}
\right]   \nonumber\\
\nu^2
&=&
\frac{m_{H0}^2}{2}
-\frac{\phi_0^2}{64\pi^2}
\left\{
\frac{3}{2}g_2^4
+\frac{3}{4}(g_1^2+g_2^2)^2
-12g_Y^4
\right\}\\
M_{W0}^2
&=&
\frac{g_2^2 \phi_0^2}{4},
M_{Z0}^2
=
\frac{(g_2^2+g_1^2) \phi_0^2}{4},\nonumber\\
m_{t0}^2
&=&
\frac{g_Y^2 \phi_0^2}{2},
\phi_0=246 \ {\rm (GeV)}\nonumber
\end{eqnarray}
Radiative corrections at the one-loop order are included in the
equations for $\nu^2$ and $\lambda$ since they are large, especially
in the case where the Higgs Boson mass is small.  The effective potential
Eq.(\ref{ini}) does not depend on $\bar\mu$ using $\lambda$ in
Eq.(\ref{para}) in this order.  We fix the masses of the Weak Bosons
and the Z-Boson as $M_{W0}=80$ (GeV) and $M_{Z0}=92$ (GeV) below.

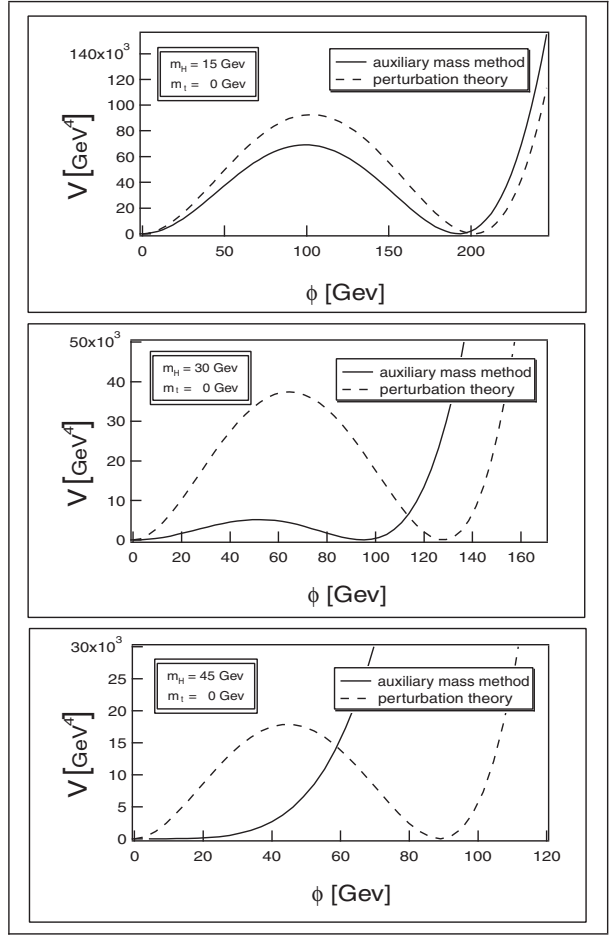
\begin{figure}
\unitlength=1cm
\begin{picture}(14,6)
\unitlength=1mm
\centerline{
\epsfxsize=9cm
\epsfbox{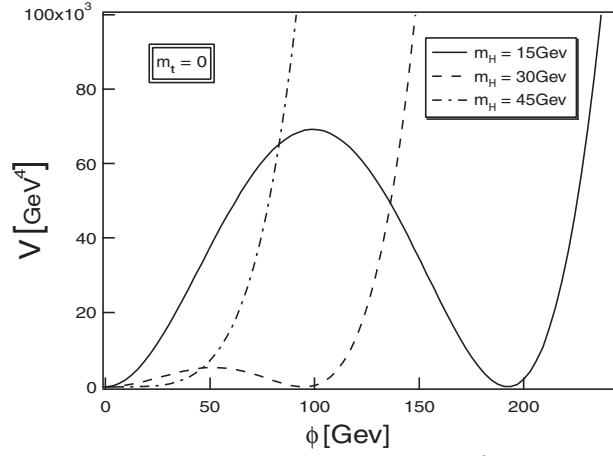} 
} 
\end{picture}
\caption{The effective potentials at the critical temperature for
$m_H=$15, 30, 45 (GeV).  The first order phase transition becomes
weaker for smaller values of the Higgs mass and disappears finally.}
\label{cripot1}
\end{figure}

\begin{figure}
\unitlength=1cm
\begin{picture}(10,6)
\unitlength=1mm
\centerline{
\epsfxsize=8cm
\epsfbox{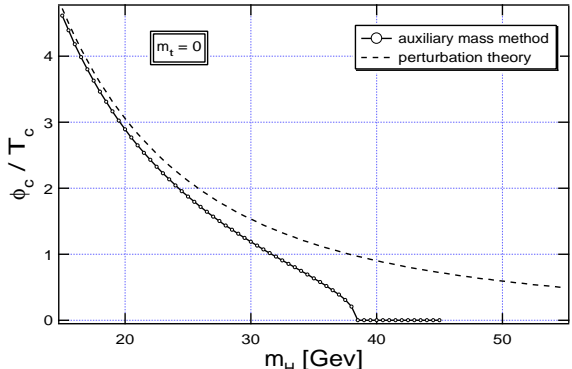} 
} 
\end{picture}
\caption{The ratio of the critical field expectation values to the
critical temperature, $\phi_c/T_c$.  The results obtained by auxiliary
mass method and the perturbation theory are similar for smaller values
of $m_H$ and different for larger values, $m_H \protect\gtrsim 30$
(GeV.}
\label{sp}
\end{figure}

We first investigate a $SU(2)\times U(1)$ gauge plus Higgs theory,
corresponding to the case $m_t=0$.  We show results obtained by setting
$M=T$ since similar results were obtained by setting $M=\frac{T}{2}$
and $M=2T$ as in the case of \cite{Ina}.  This is quite natural since
the ristriction on $M$ is $M\gtrsim T$.  The effective potentials
at the critical temperature are shown in Fig.\ref{cripot1} for
$m_H=$15, 30, 45 (GeV), respectively.  The first order phase
transition becomes weaker for smaller values of the Higgs mass and
disappears finally.  They are compared to effective potentials
obtained by the ring resumed perturbation theory at the one-loop order
without the high temperature expansion in Fig.\ref{cripot2}.  We
find clearly that they are similar for smaller values of $m_H$ and
different for larger values of $m_H$.  This is consistent with the fact
that the ring resumed perturbation theory is reliable only for smaller
values of the Higgs mass $m_H\ll M_W$\cite{Arn}.  We plot a ratio of the
critical field expectation values to the critical temperature,
$\phi_c/T_c$, as a function of $m_H$ in Fig.\ref{sp}.  This quantity
indicates the strength of the first order phase transition and important
in estimating the sphaleron rate, which plays a very important role in
the Electroweak Baryogenesis\cite{Man,Man2}.  The end-point is
determined as $m_{H,c}=38$ (GeV) from Fig.\ref{sp}. This figure also
shows that the results obtained by the auxiliary mass method and the
perturbation theory is similar for smaller values of $m_H$ and
different for larger values, $m_H \gtrsim 30$ (GeV).
\begin{figure}
\unitlength=1cm
\begin{picture}(10,12)
\unitlength=1mm
\centerline{
\epsfxsize=8cm
\epsfbox{jpcom.EPSF} 
} 
\end{picture}
\caption{The effective potentials at the critical temperature obtained 
by the auxiliary mass method and the perturbation theory for $m_H=$15,
30, 45 (GeV). They are similar for smaller values of $m_H$ and
different for larger values of $m_H$. }
\label{cripot2}
\end{figure}

\begin{figure}
\unitlength=1cm
\begin{picture}(10,6)
\unitlength=1mm
\centerline{
\epsfxsize=8cm
\epsfbox{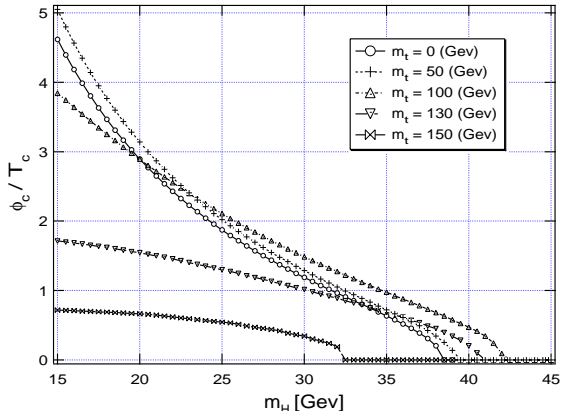} 
} 
\end{picture}
\caption{The ratio of the critical field expectation values to the
critical temperature, $\phi_c/T_c$.  The results obtained by the auxiliary
mass method and the perturbation theory are similar for smaller values
of $m_H$ and different for larger values, $m_H \protect\gtrsim 30$
(GeV).}
\label{sp2}
\end{figure}

\begin{figure}
\unitlength=1cm
\begin{picture}(10,6)
\unitlength=1mm
\centerline{
\epsfxsize=8cm
\epsfbox{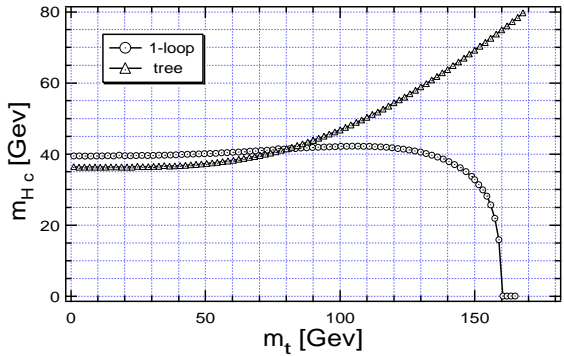} 
} 
\end{picture}
\caption{The end-points as a function of $m_t$.  The graphs labeled as 
``1-loop'' and ``tree'' are obtained with and without the
zero-temperature radiative corrections from the top quark and gauge
fields, respectively.  The end-point vanishes
 for $m_t\protect\gtrsim 160$ (GeV) in the ``1-loop'' result. }
\label{end}
\end{figure}

We next investigate more realistic cases in which the top quark mass is
finite.  The same ratios are shown in Fig.\ref{sp2} for various
values of $m_t$.  This figure shows that the strengths of the first
order phase transition are almost same for $m_t\lesssim 100$ (GeV) and
become weaker for $m_t\gtrsim 100$ (GeV) rapidly.  The end-points are
then shown in Fig.\ref{end} as a function of $m_t$. The graph
labeled ``1-loop'' is obtained using Eq.(\ref{ini}) and Eq.(\ref{para}),
which take into account the zero-temperature radiative corrections
from the top quark and gauge fields.  The contribution from the top
quark is much larger than that of the gauge fields.  On the other
hand, the graph labeled as ``tree'' is obtained without the
zero-temperature radiative correction, omitting the contributions from
$f_{G0}$ and $f_{F0}$ from Eq.(\ref{ini}) and leaving only the first
terms of Eq.(\ref{para}) for $\lambda$ and $\nu^2$.  They are not much
different for smaller values of the top quark mass, $m_t\lesssim 100$
(GeV).  Their behavior, however, differs drastically for
larger values of the top quark mass, $m_t\gtrsim 100$ (GeV).
Surprisingly, the end-point vanishes for $m_t\gtrsim 160$ (GeV) in the
``1-loop'' results though it increases in the ``tree'' results.  These
results tell us that fermionic degrees of freedom play significant
roles in the phase transition through the renormalization effects at
the zero-temperature.  We also conclude that there are no the first
order phase transitions for $m_t= 175$ (GeV), no matter how small the
Higgs Boson mass.

In the present paper, we have calculated the effective potentials of
the standard model using the auxiliary mass method at a finite
temperature.  We first investigated a $SU(2)\times U(1)$ gauge plus
Higgs theory, corresponding to the case $m_t=0$.  The phase transition
was of the first order and similar to the results obtained by the
perturbation theory for smaller $m_H\sim 15$ (GeV).  The phase
transition became weaker for larger $M_H\sim 30$ (GeV) and finally
disappeared in contrast to the results from perturbation theory.  We
found that the end-point is at $m_{H,c}=38$ (GeV) in this case.  This
is consistent with the results of the Lattice Monte Carlo simulation
\cite{Cas,Ilg,Kar,Aok} and the other non-perturbative
methods\cite{Buc,Tet,Hub} qualitatively.  The value of the end-point,
however, was smaller than those by these methods.  This may be caused
by the approximations, used to construct the evolution equation
(\ref{evo}), or used in the other papers.  The two loop effect from
the gauge fields may shift our results due to slow convergence of the
perturbation theory.  We next investigated the more realistic case in
which the top quark mass is finite.  We found that the end-point was
strongly dependent on $m_t$ and disappeared for $m_t\gtrsim 160$
(GeV).  The renormalization effects from the top quark were
significant.  Lattice Monte Carlo simulations, however, do not follow
this behavior\cite{Rum}.  We think of two possible reasons:(1)Since
our results differ from that of the Lattice Monte Carlo simulation by
factors of 2 in a $SU(2)\times U(1)$ gauge plus Higgs theory
quantitatively, the similar behavior may be found at a larger top
quark mass in the Lattice Monte Carlo simulation.(2)Since the one-loop
correction to the {\it effective potential} at the zero-temperature is
significant, the 3D effective theory, which has no Fermionic degrees
of freedom, may not reflect the effect appropriately.

Finally, the strongly first order phase transition necessary for the
Electroweak Baryogenesis was not found in the Standard Model.  We will
apply this method to extensions of the Standard Model.

The authors are supported by JSPS fellowship.

\end{document}